\newcommand{\bea}{\begin{eqnarray}}
\newcommand{\eea}{\end{eqnarray}}
\newcommand{\bear}{\begin{eqnarray*}}
\newcommand{\eear}{\end{eqnarray*}}

\documentstyle[aps,preprint]{revtex}
\begin{document}

\draft

\title
{New integrable version of the degenerate supersymmetric t-J model }

\author{
F. C. Alcaraz$^1$ and R. Z. Bariev$^{1,2}$}

\address{$^1$Departamento de F\'{\i}sica, 
Universidade Federal de S\~ao Carlos, 13565-905, S\~ao Carlos, SP
Brazil}

\address{$^2$The Kazan Physico-Technical Institute of the Russian Academy of Sciences, 
Kazan 420029, Russia}

\maketitle

\begin{abstract}
A new integrable version of the degenerate supersymmetric t-J model
is proposed. In this formulation instead of restricting single occupancy 
of electrons at each lattice site we may have up to two electrons at 
each site.  As a requirement  of exact integrability 
the hopping interaction turns out to be 
correlated with the density of electrons in the neighboring sites. 
The exact solution of the model is obtained through the coordinate Bethe 
anzatz.
\end{abstract}


\narrowtext       
Recently there has been considerable interest        
in studying low-dimensional electronic models        
with strong correlation due to the possibility that the normal        
state of the two-dimensional novel superconductivity may        
share some interesting features of a one-dimensional        
interacting electron system. Exactly solvable models of        
strongly correlated electrons which are  generalizations         
of the Hubbard and t-J models have been formulated and       
investigated [1 - 14]. It is noteworthy that some of these        
models [7,8,10] have superconducting behavior with         
 dominant superconducting correlation functions.       
       
The t-J model is a lattice model on the restricted        
electronic Hilbert space, where the occurrence of two        
electrons on the same lattice site is forbidden. The        
Hamiltonian       
of the extended version of the t-J model with the        
generalized spin $S=(N-1)/2$ has the form [15,16,10]       
\bea       
 H &=& - t\sum_{j=1}^L P_1\lbrace       
\sum_{\alpha =1}^{N}\left(c_{j,\alpha}^+       
c_{j+1,\alpha} +        
c_{j+1,\alpha}^+ c_{j,\alpha}\right)  \nonumber\\       
&+& J\sum_{\alpha\ne\beta}^{N}        
(c_{j,\alpha}^+ c_{j,\beta}        
c_{j+1,\beta}^+ c_{j+1,\alpha} -       
n_{j,\alpha}n_{j+1,\beta})        
\rbrace P_1       
\eea 
where $c_{j,\alpha}$ and $n{j,\alpha} = c_{j,\alpha}^+c_{j,\alpha}$ 
are the standard fermionic and density operators and $P_1$ projects out
from the Hilbert space the states with more than single occupancy at a
given site.     
The model (1) is not integrable for all the values        
of the parameter $J/t$ and arbitrary       
band-filling, but only at the supersymmetric        
point $J=t$.  By construction the spin degrees of freedom  of       
 model (1) form a $SU(N)$-invariant subspace. 

In this communication we introduce a new        
version of the supersymmetric t-J model.       
In this version instead of the operator $ P_1$       
we introduce the operator $ P_2$       
which is a projection operator into the        
subspace where at most double occupancy        
is allowed at the sites. Differently from (1) instead of N allowed states we 
have in this case $(N^2+N+2)/2$       
possible states at a given site $j$, namely       
\bea       
|0>; \hspace{1cm}c_{j,\alpha}^+|0> (\alpha =1,2,...,N); \nonumber\\       
c_{j,\alpha}^+c_{j,\beta}^+|0> (\alpha\ne\beta  =1,2,...,N).       
\eea       
In a periodic lattice of length L the        
Hamiltonian of our model reads       
\bea       
H&=& -\sum_{j=1}^L P_2 H_{j,j+1}P_2,\nonumber\\       
H_{j,k} &=& \sum_{\alpha}(c_{j,\alpha}^+c_{k,\alpha}+h.c.)\nonumber\\       
&\times& \lbrace\mbox{exp}\lbrack \frac{1}{2}\eta       
\sum_{\beta(\ne\alpha)}(n_{j\beta}+n_{k,\beta})\rbrack        
-\sum_{\beta\ne\gamma(\ne\alpha)}n_{j,\beta}n_{k,\gamma}\rbrace\nonumber\\       
&+&\sum_{\alpha<\beta}\lbrack U(n_{j,\alpha}n_{j,\beta} +        
n_{k,\alpha}n_{k,\beta})       
+t_p(c_{j,\alpha}^+c_{j,\beta}^+c_{k,\alpha}c_{k,\beta}+h.c.)\rbrack        
\nonumber\\       
&-& J\sum_{\alpha,\beta}c_{j,\alpha}^+c_{k,\beta}^+c_{j,\beta}c_{k,\alpha}       
\lbrack\sum_{\gamma\ne(\alpha,\beta)}(n_{j,\gamma}+n_{k,\gamma}) -       
\sum_{\gamma\ne\delta\ne(\alpha,\beta)}n_{j,\gamma}n_{k,\delta}\rbrack       
\nonumber\\       
&-& \sum_{\alpha<\beta\ne\gamma}       
(c_{j,\alpha}^+c_{j,\beta}^+c_{k,\gamma}^+c_{j,\gamma}c_{k,\beta}c_{j,\alpha}       
+h.c.)\nonumber\\       
&+& J\sum_{\alpha<\beta\ne\gamma<\delta}       
c_{j,\alpha}^+c_{j,\beta}^+c_{k,\gamma}^+c_{k,\delta}^+       
c_{j,\delta}c_{j,\gamma}c_{k,\beta}c_{k,\alpha}\nonumber\\       
&-& J\sum_{\alpha<\beta\ne\gamma}(n_{j,\alpha}n_{j,\beta}n_{k,\gamma}       
+n_{k,\alpha}n_{k,\beta}n_{j,\gamma})\nonumber\\       
&+& J\sum_{\alpha<\beta\ne\gamma<\delta}n_{j,\alpha}n_{j,\beta}       
n_{k,\gamma}n_{k,\delta}.       
\eea       
Similarly as in the standard t-J model we have established        
the integrability of (1) only at a special point       
\bea       
t_p = J = U =\varepsilon (e^{\eta} - 1),  \mbox{at }       
   \eta = \ln 2, \varepsilon =\pm 1       
\eea       
It is interesting to point that in the case $N = 2$, differently from (1),
the model (3) is integrable at an arbitrary parameter $\eta$ [12 - 14].       
       
The model (1) has been constructed in the following way. We consider        
all possible interactions that conserve the number        
of particles of each species separately and satisfy
the constraint imposed by        
$P_2$ that not more than double occupancy is allowed at each site.       
Moreover we impose for the amplitude of the eigenfunctions of the       
Hamiltonian (1) in the sector with $n_{\alpha}$ particles of       
species $\alpha (\alpha = 1,2,...,N)$ the following ansatz       
\bea       
\Psi (x_1,\alpha_1;...;x_n,\alpha_n) =        
\sum_{P}A_{p_1...p_n}^{\alpha_{Q_1}...\alpha_{Q_n}}        
\prod_{j=1}^{n}\mbox{exp}(ik_{P_j}x_{Q_j})       
\eea       
where $Q = (Q_1,...,Q_n)$ is the permutation of the $N$ particles        
such that their coordinates satisfy 
$ 1 \le x_{Q_1}\le x_{Q_2}\le ... \le x_{Q_n} \le L$.       
The sum is over all permutations $P = (P_1,...,P_n)$       
of integers $1,2,...,n$. In the case we have two particles        
at the same site $x_{Q_l }= x_{Q_{l+1}}$ the ansatz (5) is replaced by       
\bea       
\Psi (x_1,\alpha_1;...;x_n,\alpha_n) =        
\sum_{P}A_{p_1...p_lp_{l+1}...p_n}^       
{\alpha_{Q_1}...\overline{\alpha_{Q_l}\alpha_{Q_{l+1}}}       
...\alpha_{Q_n}}        
\prod_{j=1}^{n}\mbox{exp}(ik_{P_j}x_{Q_j})       ,
\eea       
where the bar at the $l^{th}$ and  $(l+1)^{th}$ position of the        
superscript indicates the pair position.       
Secondly we consider the model (3) in the sector where we have only       
two electrons. In this case the projector $P_2$ does not produce        
any restrictions and our problem is exactly equivalent to that       
of a model with only two species. The coefficients       
$A_{p_1...p_n}^{\alpha_{Q_1}...\alpha_{Q_n}}$ arising        
from the different permutation ${Q}$ are connected with each        
other by the elements of the two-particle S-matrix       
\bea       
A_{...p_1P_2...}^{...\alpha\beta...} =        
\sum_{\delta,\gamma = 1}^N        
S_{\alpha\beta}^{\gamma\delta}(k_{p_1},k_{p_2})       
A_{...p_2p_1...}^{...\delta\gamma...}      . 
\eea       
A necessary condition of the compatibility of equations (7)       
is the fulfillment of the Yang-Baxter equations [17,18]. There are       
two integrable models with $N=2$ and their S-matrices have       
such factorizable form. These models are the Hubbard model       
[1] and the correlated hopping model [6,7,12-14]. We may try to use as       
the fundamental building block of a general S-matrix of our model       
(3) the two-particle scattering matrix of these models. Here       
we use the second possibility and choose the S-matrix as in        
the correlated hopping model[12 - 14]       
\bea       
S_{1,2}(k_1,k_2) = \frac{\lambda_1 - \lambda_2 - iP_{12}}       
{\lambda_1- \lambda_2 - i},       
\eea       
where the operator $P_{12}$ interchanges the species variables        
${\alpha_1}$ and ${\alpha_2}$ and       
\bea       
\lambda_j = \lambda(k_j) = \cases{       
\cot\frac{1}{2}k,& ${\varepsilon}=-1$\cr       
-\tan\frac{1}{2}k,& ${\varepsilon}=+1$\cr}   .    
\eea       
       
To complete the proof of the Bethe anzatz        
(5-6) we must consider the eigenvalue equations        
 in the sector where the total number of particles $n = 3,4$. 
This gives a        
complicated system of equations for the        
coupling constants of the Hamiltonian (1).       
The solution of this system is presented in the
Hamiltonian (3).       
The periodic boundary condition for the system       
on the finite interval (0,L) gives us the Bethe ansatz        
equation. In order to obtain these equations we must        
diagonalize the transfer matrix of a related        
inhomogeneous vertex model with non-intersecting        
strings [19 - 20]. The Bethe ansatz equations are written        
in terms of the charge rapidities $\lambda^{(0)}$ and additional 
$N-1$-spin rapidities $\lambda^{(l)} (l=1,\ldots,N-1)$       
\bea       
\left(\frac{\lambda_j^{(0)}-i}{\lambda_j^{(0)}+i}\right)^L       
&=&\prod_{\alpha = 1}^{M_1}\frac       
{\lambda_j^{(0)}-\lambda_{\alpha}^{(1)}-i/2}       
{\lambda_j^{(0)}-\lambda_{\alpha}^{(1)}+i/2}\nonumber\\       
\prod_{\beta =1}^{M_l}\frac{\lambda_{\alpha}^{(l)} -\lambda_{\beta}^{(l)} -i}       
{\lambda_{\alpha}^{(l)} -\lambda_{\beta}^{(l)} +i} &=& -       
\prod_{\beta =1}^{M_{l-1}}\frac       
{\lambda_{\alpha}^{(l)} -\lambda_{\beta}^{(l-1)} -i/2}       
{\lambda_{\alpha}^{(l)} -\lambda_{\beta}^{(l-1)} +i/2}       
\prod_{\beta =1}^{M_{l+1}}\frac       
{\lambda_{\alpha}^{(l)} -\lambda_{\beta}^{(l+1)} -i/2}       
{\lambda_{\alpha}^{(l)} -\lambda_{\beta}^{(l+1)} +i/2}       
\eea       
\bea       
l = 1,...,N-1, M_0 =n, M_N = 0, \alpha = 1,...,M_l,        
\lambda_j^{(0)}=\lambda_j,\nonumber       
\eea       
where $n_j = M_{j-1} - M_j $ is the number of particles       
of species $j$.       
The eigenenergies  of the system are given in terms of the        
 solutions $\{\lambda_{\alpha}^{(k)}\}$ and are  given by     
\bea       
E = -2\sum_{j=1}^n \cos k_j =        
2\varepsilon(n- \sum_{j=1}^n\frac{2}{(\lambda_j^{(0)})^2 + 1}).       
\eea 
The corresponding 
 magnetization of the eigenstate is given by  
a pure Zeeman splitting of the $SU(N)$
spin multiplet 
\bea
S^z = \frac{1}{2}(N-1)n - \sum_{l=1}^{N-1}M_l.
\eea 
The ground state and  excitations of the system are given by 
inserting in (11) the 
solutions of equations (10). The rapidities have,
in general, complex values and their classification is analogous to  
that of the degenerate electron gas with an attractive $\delta$-function
potential[21]: (i) real charge rapidities, belonging to the set 
$\lambda_{\alpha}^{(0)}$,  and correspond to unpaired propagating 
electrons; (ii) complex spin and charge rapidities, which correspond
to bound states of electrons with different spin components; and (iii)
strings of complex spin rapidities, representing spin states. A complex 
of $m$ electrons $(m\le 0)$ is characterized [22] 
by one real $\zeta^{(m-1)}$
rapidity and the remaining ones given by
\bea
\lambda_{p_l}^{(l)} &=& \zeta^{(m-1)} + i\frac{p_l}{2}, \\
l &\le& m-1,\;\;\;\;
p_l = -(m-l-1), -(m-l-3),...,(m-l-1). \nonumber
\eea 
These spin and charge strings form the classes  (i) and (ii) above 
mentioned. In the class 
(iii) there is a set of rapidities $\{\lambda_{\alpha}^{(l)}\} 
(l =1,2,...,N-1)$, forming asymptotically strings of maximum size $n$, 
 \bea
\lambda_n^{(l)\mu} = \Lambda_n^{(l)} + i\mu/2 + \delta^{(l)\mu},\\
\mu = -(n-1), -(n-3),...,(n-1). \nonumber 
\eea
Here $\delta^{(l)\mu} = O(e^{-aL}), a> 0$, vanishes in the thermodynamic 
limit $L\to \infty$ and $\Lambda_n^{(l)}$ are real numbers. 
The above rapidities (13) and (14) are substituted into (10) and the resulting 
coupled equations for the real $\zeta^{(l)}$ and $\Lambda_n^{(l)}$ are 
logarithmized. This procedure generated a set of integer quantum numbers 
for each
set of rapidities: $\rho^{(l)}(\zeta)$ for the real $\zeta^{(l)}, 
l = 0,...,N-1$ and similarly for the corresponding "hole" functions. In the
thermodynamic limit the Bethe
ansatz equations reduce to sets of coupled linear integral equations
for the density functions. After Fourier transforming these
equations we obtain
\bea
\hat\rho_h^{(l)}(\omega) &+& \hat \rho^{(l)}(\omega) + 
\sum_{q=0}^{N-1} \hat \rho^{(q)}(\omega)\mbox{exp}\lbrack -(l + q - p_{l,q})
|\omega|/2\rbrack\nonumber\\
&\times &\frac{\sinh[(p_{l,q}+1)\omega/2]}{\sinh(\omega/2)}+
\sum_{n=1}^{\infty}\hat 
\sigma_n^{(l+1)}(\omega)\mbox{exp}(-n|\omega|/2)\nonumber\\
&=&2e^{-(l+1)|\omega|/2}\cosh\omega/2, \;\;\;l = 0,...,N-1,\\
\hat\sigma_{mh}^{(l)}(\omega &=& \hat \rho_h^{(l-1)}(\omega)\mbox{exp}
(-m|\omega|/2)\nonumber\\
&+&\sum_{n=1}^{\infty}\lbrack \hat\sigma_n^{(l-1)}(\omega) +  
\hat\sigma_n^{(l+1)}(\omega) - \cosh(\omega/2)\hat 
\sigma_n^{(l)}(\omega)\rbrack \nonumber\\
&\times&\mbox{exp}(-\mbox{max}(m,n)|\omega|/2)\frac{\sinh[\mbox{min}(m,n)
\omega/2]}
{\sinh(\omega/2)},\;\;\; l = 1,...,N-1.
\eea
The last set of equations holds for $m = 1,...,\infty$ with
$\sigma_m^{(0)}, \sigma_{mh}^{(0)},\sigma_m^{(N)}, $ and $\sigma_{mh}^{(N)}$
being identically zero, and $p_{l,q} = \mbox{min}(l,q) - \delta_{l,q}$ in 
eq. (15). These 
equations are identical to those appearing in  
the degenerate electron gas with an 
attractive $\delta$-function potential [21] and, apart from the
driving terms, are also identical to those of the degenerate 
supersymmetrical t-J model [14, 22]. 
The expression for the energy however is different and is given by
\bea
\frac{1}{L}E = 2\varepsilon\lbrace\rho - \sum_{l=0}^{N-1}\int d\zeta
\rho^{(l)}(\zeta)\lbrack\frac{(l+2)}{(l+2)^2/4 + \zeta^2} +
\frac{l}{l^2/4 + \zeta^2}\rbrack\rbrace .
\eea
Next we discuss the ground state properties. First consider 
$\varepsilon = - 1$. As in the ordinary t-J model the ground state is 
described by the bound complexes (13) with  the maximum length N. In order to 
accommodate such  complexes in the ground state we assume that 
the number of electrons n is a 
multiple of $N$. In the thermodynamical limit, $L,n\to\infty$ we obtain, from 
(15-16), 
\bea
2\pi\rho(\Lambda) &+&\lbrack \int_{-\infty}^{-v_0} + \int_{v_0}^{\infty}\rbrack\lbrack\sum_{l=1}^{N-1}
\theta'(\frac{1}{l}(\Lambda-\Lambda'))\rbrack\rho(\Lambda')d\Lambda'
= \theta'(\frac{2\Lambda}{N+1}) +  \theta'(\frac{2\Lambda}{N-1})\nonumber \\
\lbrack \int_{-\infty}^{-v_0} + \int_{v_0}^{\infty}\rbrack&\rho&(\Lambda)d\Lambda = \frac{1}{N}\rho, 
\;\;\;\;\; \rho = \frac{n}{L},
\eea
where 
\bea
\theta'(\frac{1}{l}\Lambda)= \frac{2l}{l^2 + \Lambda^2}\nonumber.
\eea
The solution $\rho(\Lambda)$ of these equations yields the ground-state 
energy per site
\bea
\frac{1}{L}E = 
-2\lbrace\rho - \lbrack \int_{-\infty}^{-v_0} + \int_{v_0}^{\infty}\rbrack
\lbrack\theta'(\frac{2\Lambda}{N+1}) + \theta'(\frac{2\Lambda}{N-1})\rbrack
\rho(\Lambda)d\Lambda\rbrace.
\eea 
In order to solve numerically the integral equations (18) it is convenient 
 to rewrite them  in the following form
\bea
\rho(\Lambda) &-& \frac{1}{2\pi}\int_{-v_0}^{v_0}K_1(\Lambda - \Lambda')
\rho(\Lambda')d\Lambda' = \frac{1}{2\pi} K_2(\Lambda),\nonumber\\
\int_{-v_0}^{v_0}\rho(\Lambda)d\Lambda &=& 2 - \rho ,
\eea
where
\bea
K_i(\Lambda) &=& \int_{-\infty}^{\infty}\tilde K_i(p)e^{-ip\Lambda}dp \;\;\; 
i=1,2 ,\nonumber\\
\tilde K_1(p) &=&\frac{e^{-|p|/2}\sinh\frac{1}{2}(N-1)p}{\sinh\frac{1}{2}Np}; 
\;\; \tilde K_2(p) =\frac{e^{-|p|/2}\sinh p}{\sinh \frac{1}{2}Np}.\nonumber
\eea
For the ground-state energy we now have
\bea
\frac{1}{L}E = -2\lbrace \rho -\Phi_2(0) + \int_{-v_0}^{v_0}K_2(-\Lambda)
\rho(\Lambda)d\Lambda\rbrace ,
\eea
where
\bea
\Phi_2(\Lambda) = \int_{-\infty}^{\infty}
(e^{-\frac{N+1}{2}|p|} + e^{-\frac{N-1}{2}|p|})K_2(p)e^{-ip\Lambda}dp .
\eea
The energy density $E/L$ (19) as a function of the density $\rho =n/L$ is 
shown in Fig. 1 
for several values of $N$. 
For the sake of comparison we show in Fig. 2 the ground-state energy of the 
ordinary degenerated t-J model, for some values of $N$. These figures show 
that for small values of the density the ground-state energy of both models 
are the same. The same similarity is also expected for all densities $\rho 
\leq 1$ but  for large values of $N$. On the other hand,  for small values 
of $N$ and large densities, where 
the site-exclusion effect is more important, these figures show 
quite distinct behavior for both models.

Secondly consider the case of ferromagnetic coupling
$\varepsilon = +1$. In contrast with 
 the ordinary t-J model the ground state is described by bound 
complexes where the  rapidities $\lambda^{(0)}$ have length 2. We have in this 
case an equal 
number of electrons with  two different spin components. The 
situation is analogous to that of the correlated hopping model [14]. 
The energy density, 
\bea
\frac{1}{N}E = 2\lbrace\rho - \int_{-v_0}^{v_0}
\lbrack\theta'(\frac{2}{3}\Lambda) + \theta'(2\Lambda)\rbrack
\rho^{(1)}(\Lambda)d\Lambda\rbrace
\eea
is now given in terms of the solution of the integral equations
\bea
2\pi\rho^{(1)}(\Lambda) &+& \int_{-v_0}^{v_0}\theta'(\Lambda - \Lambda')
\rho^{(1)}(\Lambda')d\Lambda' = \theta'(\frac{2}{3}\Lambda) + 
\theta'(2\Lambda), 
\nonumber\\
\int_{-v_0}^{v_0}\rho^{(1)}(\Lambda)d\Lambda &=& \frac{1}{2}\rho .
\eea
The energy density $E/L$, as a function of $\rho$, obtained 
from the numerical solution of (23-24), is shown in Fig. 3.

To summarise, we have presented an integrable       
$(N^2 + N +2)/2$-state version of the degenerate        
supersymmetric $t-J$ model where at most two electrons are allowed in a 
given site. We have solved the model        
by the coordinate Bethe       
ansatz method and derived the Bethe ansatz equations.       
Our results naturally raise interesting problems       
to be solved:       
the construction of the anisotropic version        
of the Hamiltonian (1) as well as its graduated version.       
In the latter case one should have an integrable model of        
interacting chains 
with spin $S =1/2$ and $1$. The isotropic version for $N=2$ of        
this model has been solved recently by Frahm et al. [23].       
Finally it is possible to construct new  generalized  
integrable models by the        
introduction of the operator $P_l (l >2)$ which is a projection        
operator into the subspace where at most l-times occupancy is allowed       
at the sites. As a fundamental model we may use in this case the        
generalization of the correlated hopping model where we have $l$ 
distinct species with      
single particle as well as multi-particle hopping.       
For $l=3$ we may use as the fundamental building block
of the general scattering matrix the S-matrix of a model which was constructed 
recently in refs. [24,25]. For an arbitrary $l$,  as a 
result of this 
construction,  we may obtain the integrable model with the symmetry of
quantum superalgebra  $U_q[gl(m+1|N-m)]_{k=l}$.
   Certainly it will be interesting to study the        
physical properties of these quantum chains particularly their 
 phase diagram and        
critical exponents.       
       
We thank  M. J. Martins and H. Frahm for useful discussions.       
This work was supported in part by Conselho Nacional        
de Desenvolvimento 
Cient\'{\i}fico e Tecnol\'ogico  - CNPq - Brazil,        
 and by The Russian Foundation of Fundamental Investigations 
under Grant No 99-02-17646.

\newpage
\Large
\begin{center}
Figure Captions
\end{center}
\normalsize
\vspace{1cm}
 
\noindent Figure 1 - Ground-state energy per site $E/L$ as a function 
of the density $\rho=n/L$ for the exact integrable model (3) with 
$t_p=J=U=\epsilon  = 1$ and some values of $N$.
\vspace{1cm}
 
\noindent Figure 2 - Ground-state energy per site $E/L$ as a function 
of the density $\rho=n/L$ for the degenerated t-J model (1) at 
$J=t=1$,  
for  some values of $N$.
\vspace{1cm}
 
\noindent Figure 3 - Ground-state energy per site $E/L$ as a function 
of the density $\rho=n/L$ for the exact integrable model (3) with 
$t_p=J=U=\epsilon  = -1$.

\vspace{2cm}

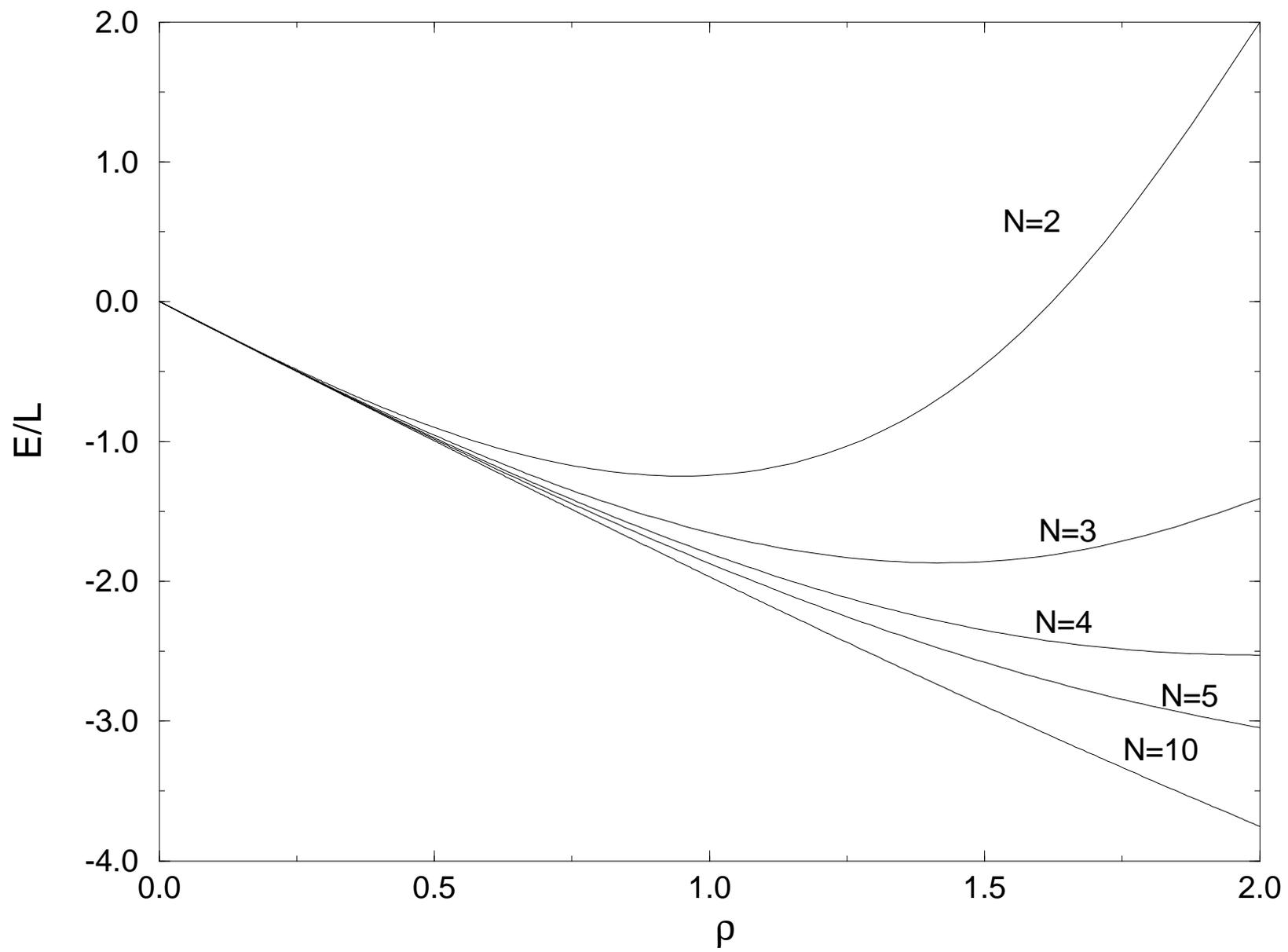
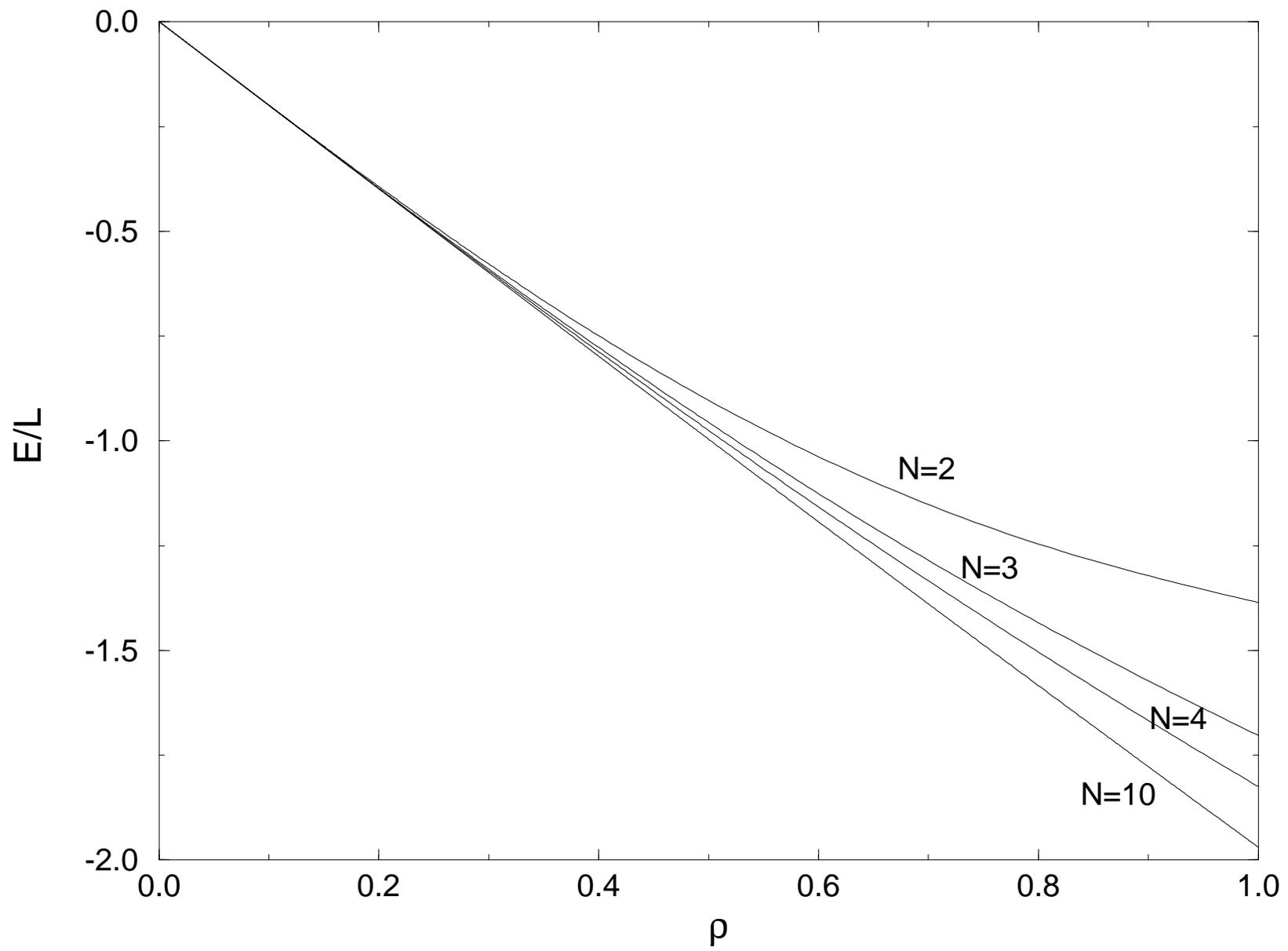
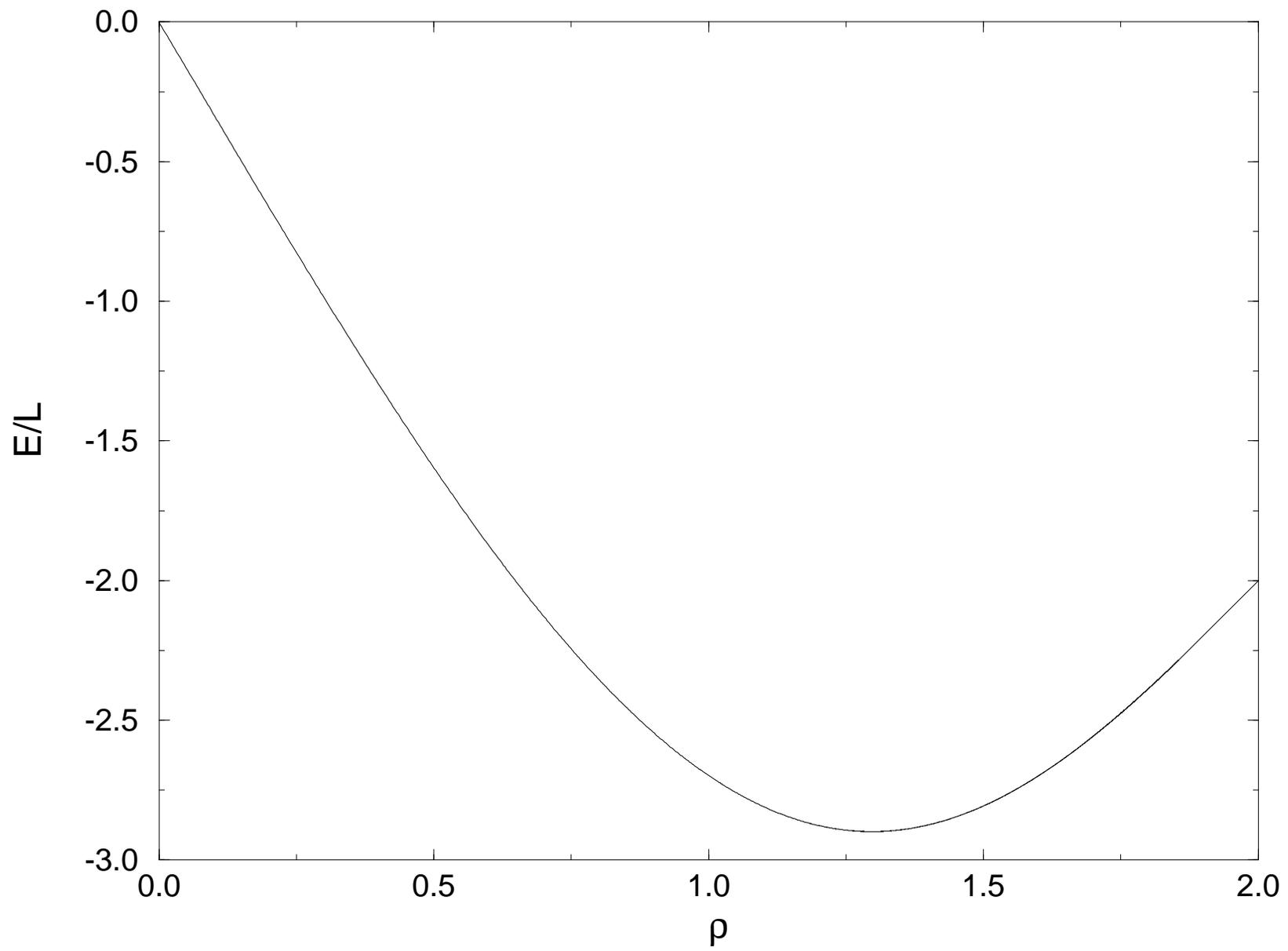
\end{document}